\documentclass[twocolumn,showpacs,aps,prl,superscriptaddress]{revtex4}

\usepackage{graphicx}
\usepackage{dcolumn}
\usepackage{amsmath}
\usepackage{epsfig}
\usepackage{color}
\usepackage{colordvi}
\usepackage{amsfonts}            
\usepackage{fancyhdr,lineno,multirow,rotating} 

\RequirePackage{xspace}

\usepackage{relsize}
\def\babar{\mbox{\slshape B\kern-0.1em{\smaller A}\kern-0.1em
    B\kern-0.1em{\smaller A\kern-0.2em R}}}

\def\epem       {\ensuremath{e^+e^-}\xspace}

\def\ccbar {\ensuremath{c\overline c}\xspace}

\def\bbbar {\ensuremath{b\overline b}\xspace}

\def\piz   {\ensuremath{\pi^0}\xspace}

\def\pip   {\ensuremath{\pi^+}\xspace}
\def\pim   {\ensuremath{\pi^-}\xspace}

\def\Kbar  {\kern 0.2em\overline{\kern -0.2em K}{}\xspace}

\def\Kz    {\ensuremath{K^0}\xspace}
\def\Kzb   {\ensuremath{\Kbar^0}\xspace}
\def\KzKzb {\ensuremath{\Kz \kern -0.16em \Kzb}\xspace}
\def\Kp    {\ensuremath{K^+}\xspace}
\def\Km    {\ensuremath{K^-}\xspace}

\def\KpKm  {\ensuremath{\Kp \kern -0.16em \Km}\xspace}

\def\Dbar    {\kern 0.2em\overline{\kern -0.2em D}{}\xspace}

\def\Dz      {\ensuremath{D^0}\xspace}
\def\Dzb     {\ensuremath{\Dbar^0}\xspace}
\def\DzDzb   {\ensuremath{\Dz {\kern -0.16em \Dzb}}\xspace}
\def\Dp      {\ensuremath{D^+}\xspace}
\def\Dm      {\ensuremath{D^-}\xspace}

\def\DpDm    {\ensuremath{\Dp {\kern -0.16em \Dm}}\xspace}
\def\Dstar   {\ensuremath{D^*}\xspace}

\def\Dstarp  {\ensuremath{D^{*+}}\xspace}

\def\B       {\ensuremath{B}\xspace}
\def\Bbar    {\kern 0.18em\overline{\kern -0.18em B}{}\xspace}

\def\Bz      {\ensuremath{B^0}\xspace}
\def\Bzb     {\ensuremath{\Bbar^0}\xspace}
\def\BzBzb   {\ensuremath{\Bz {\kern -0.16em \Bzb}}\xspace}
\def\Bu      {\ensuremath{B^+}\xspace}
\def\Bub     {\ensuremath{B^-}\xspace}

\def\BpBm    {\ensuremath{\Bu {\kern -0.16em \Bub}}\xspace}

\mathchardef\Upsilon="7107
\def\Y#1S{\ensuremath{\Upsilon{(#1S)}}\xspace}

\mathchardef\Deltares="7101
\mathchardef\Xi="7104
\mathchardef\Lambda="7103
\mathchardef\Sigma="7106
\mathchardef\Omega="710A

\def\Deltabar{\kern 0.25em\overline{\kern -0.25em \Deltares}{}\xspace}
\def\Lbar{\kern 0.2em\overline{\kern -0.2em\Lambda\kern 0.05em}\kern-0.05em{}\xspace}
\def\Sigbar{\kern 0.2em\overline{\kern -0.2em \Sigma}{}\xspace}
\def\Xibar{\kern 0.2em\overline{\kern -0.2em \Xi}{}\xspace}
\def\Obar{\kern 0.2em\overline{\kern -0.2em \Omega}{}\xspace}
\def\Nbar{\kern 0.2em\overline{\kern -0.2em N}{}\xspace}
\def\Xb{\kern 0.2em\overline{\kern -0.2em X}{}\xspace}

\newcommand{\tev}{\ensuremath{\mathrm{\,Te\kern -0.1em V}}\xspace}
\newcommand{\gev}{\ensuremath{\mathrm{\,Ge\kern -0.1em V}}\xspace}
\newcommand{\mev}{\ensuremath{\mathrm{\,Me\kern -0.1em V}}\xspace}
\newcommand{\kev}{\ensuremath{\mathrm{\,ke\kern -0.1em V}}\xspace}
\newcommand{\ev}{\ensuremath{\mathrm{\,e\kern -0.1em V}}\xspace}
\newcommand{\gevc}{\ensuremath{{\mathrm{\,Ge\kern -0.1em V\!/}c}}\xspace}
\newcommand{\mevc}{\ensuremath{{\mathrm{\,Me\kern -0.1em V\!/}c}}\xspace}
\newcommand{\gevcc}{\ensuremath{{\mathrm{\,Ge\kern -0.1em V\!/}c^2}}\xspace}
\newcommand{\mevcc}{\ensuremath{{\mathrm{\,Me\kern -0.1em V\!/}c^2}}\xspace}

\def\invfb   {\ensuremath{\mbox{\,fb}^{-1}}\xspace}

\def\mus  {\ensuremath{\rm \,\mus}\xspace}

\def\mus        {\ensuremath{\,\mu{\rm s}}\xspace}    

\def\to                 {\ensuremath{\rightarrow}\xspace}

\def\pep2{PEP-II}

\def\gsim{{~\raise.15em\hbox{$>$}\kern-.85em
          \lower.35em\hbox{$\sim$}~}\xspace}
\def\lsim{{~\raise.15em\hbox{$<$}\kern-.85em
          \lower.35em\hbox{$\sim$}~}\xspace}

\xspace

\newcommand{\nimBaseC}       {Nucl.\ Instr.\ and Methods\xspace}

\newcommand{\nim}       [1]  {\nimBaseC~{\bf #1}}

\def\jetset74   {\mbox{\tt Jetset \hspace{-0.5em}7.\hspace{-0.2em}4}\xspace}

\long\def\inst#1{\par\nobreak\kern 4pt\nobreak
    {\it #1}\par\vskip 10pt plus 3pt minus 3pt}

\def\figurebox#1#2#3{%
    \def\arg{#3}%
    \ifx\arg\empty
    {\hfill\vbox{\hsize#2\hrule\hbox to #2{\vrule\hfill\vbox to #1{\hsize#2\vfill}\vrule}\hrule}\hfill}%
    \else
    {\hfill\epsfbox{#3}\hfill}%
    \fi}

\newcommand{\DstarDzpis}{\ensuremath{\Dstarp \to \Dz \pip_s}}
\newcommand{\Dzpppz}{\ensuremath{\Dz \to \pim\pip\piz}}
\newcommand{\Dzpp}{\ensuremath{\Dz \to \pim\pip}}
\newcommand{\Dzkppz}{\ensuremath{\Dz \to K^-\pip\piz}}
\newcommand{\Dzkkpz}{\ensuremath{\Dz \to K^{-}K^{+}\piz}}
\newcommand{\Dzkk}{\ensuremath{\Dz \to K^{-}K^{+}}}

\newcommand{\pmpppz}{\ensuremath{\pi^- \pi^+ \pi^0}}
\newcommand{\kmkppz}{\ensuremath{K^- K^+ \pi^0}}

\newcommand{\Deltam}{\ensuremath{\Delta m}}

\newcommand{\pppz}{\ensuremath{\pim\pip\piz}}
\newcommand{\kppz}{\ensuremath{K^-\pip\piz}}
\newcommand{\kkpz}{\ensuremath{K^-K^+\piz}}

\begin{document}

\leftline{\babar\  PUB 06/048}
\leftline{SLAC  PUB 11977}

\title{
{\large \bf \boldmath
Precise Branching Ratio Measurements of the Decays \Dzpppz\, and \Dzkkpz}
}

%
\author{B.~Aubert}
\author{R.~Barate}
\author{M.~Bona}
\author{D.~Boutigny}
\author{F.~Couderc}
\author{Y.~Karyotakis}
\author{J.~P.~Lees}
\author{V.~Poireau}
\author{V.~Tisserand}
\author{A.~Zghiche}
\affiliation{Laboratoire de Physique des Particules, F-74941 Annecy-le-Vieux, France }
% ...
\collaboration{The \babar\ Collaboration}
\noaffiliation
\date{\today}

\begin{abstract}
Using 232 fb${}^{-1}$ of $e^+e^-$ collision data recorded by the \babar\ 
experiment, we measure the rates of three-body Cabibbo-suppressed decays 
of the \Dz\ meson relative to the Cabibbo-favored decay:
$\frac{ \B(\Dzpppz)}{ \B(\Dzkppz)} = 
(10.59 \pm 0.06 \pm 0.13) \times 10^{-2} $ and 
$\frac{ \B(\Dzkkpz)}{ \B(\Dzkppz)} = 
(2.37 \pm 0.03 \pm 0.04) \times 10^{-2} $, where the errors are statistical 
and systematic, respectively. 
The precisions of these measurements are significantly
better than those of the current world average values.
\end{abstract}


\maketitle

\begin{figure}[!htbp]
\includegraphics[height=1.6in, width=2.8in]{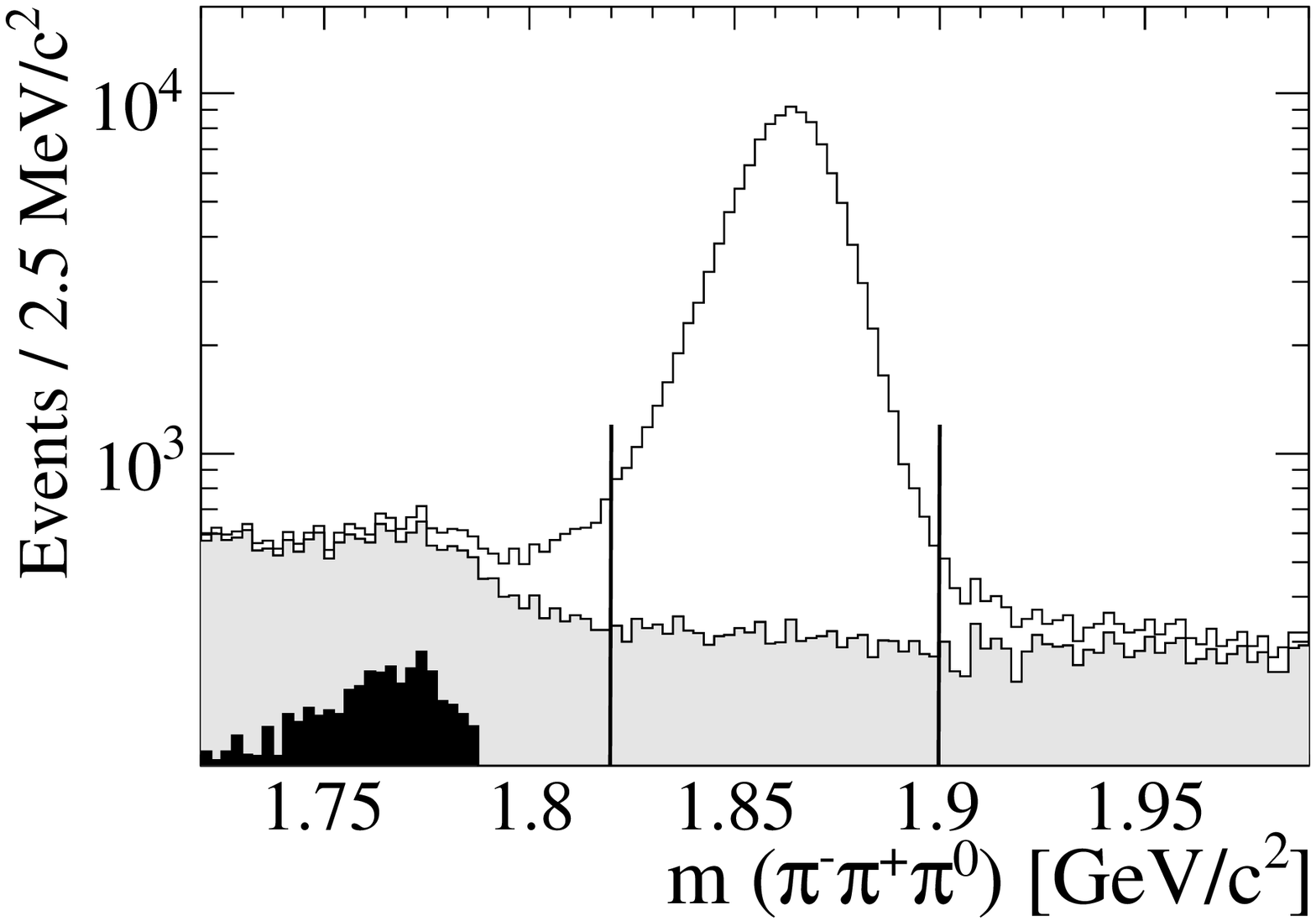}
\includegraphics[height=1.6in, width=2.8in]{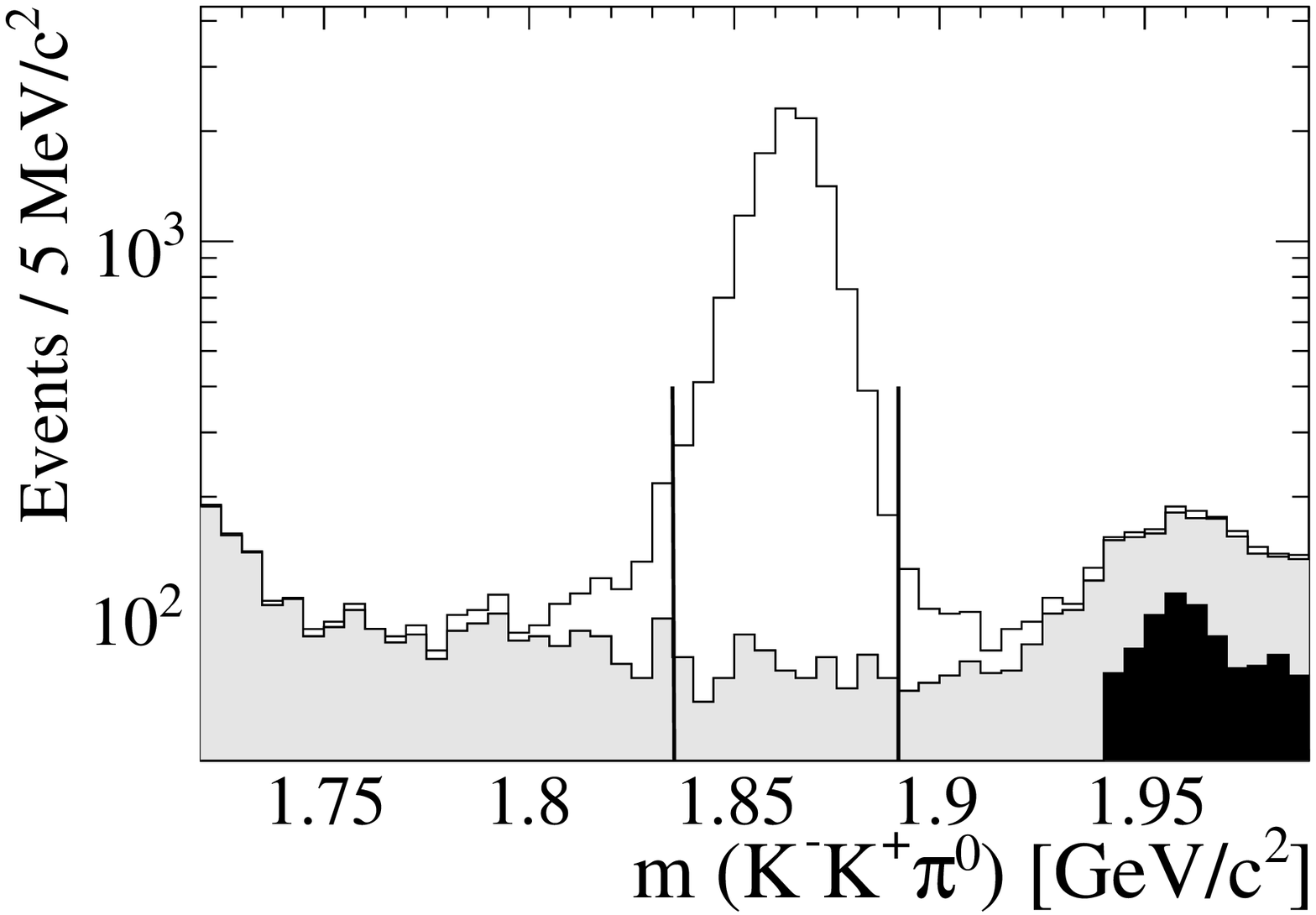}
\caption{Simulated \pmpppz\ (top) and \kmkppz\ (bottom) invariant mass distributions.
Signal events are shown as open histograms. Combinatorial and reflection 
backgrounds are shown by the light and dark shaded histograms, respectively.
The signal region is delimited by the vertical lines.}
\label{generic}
\end{figure}

\begin{figure}[!htbp]
\includegraphics[height=1.8in, width=2.8in]{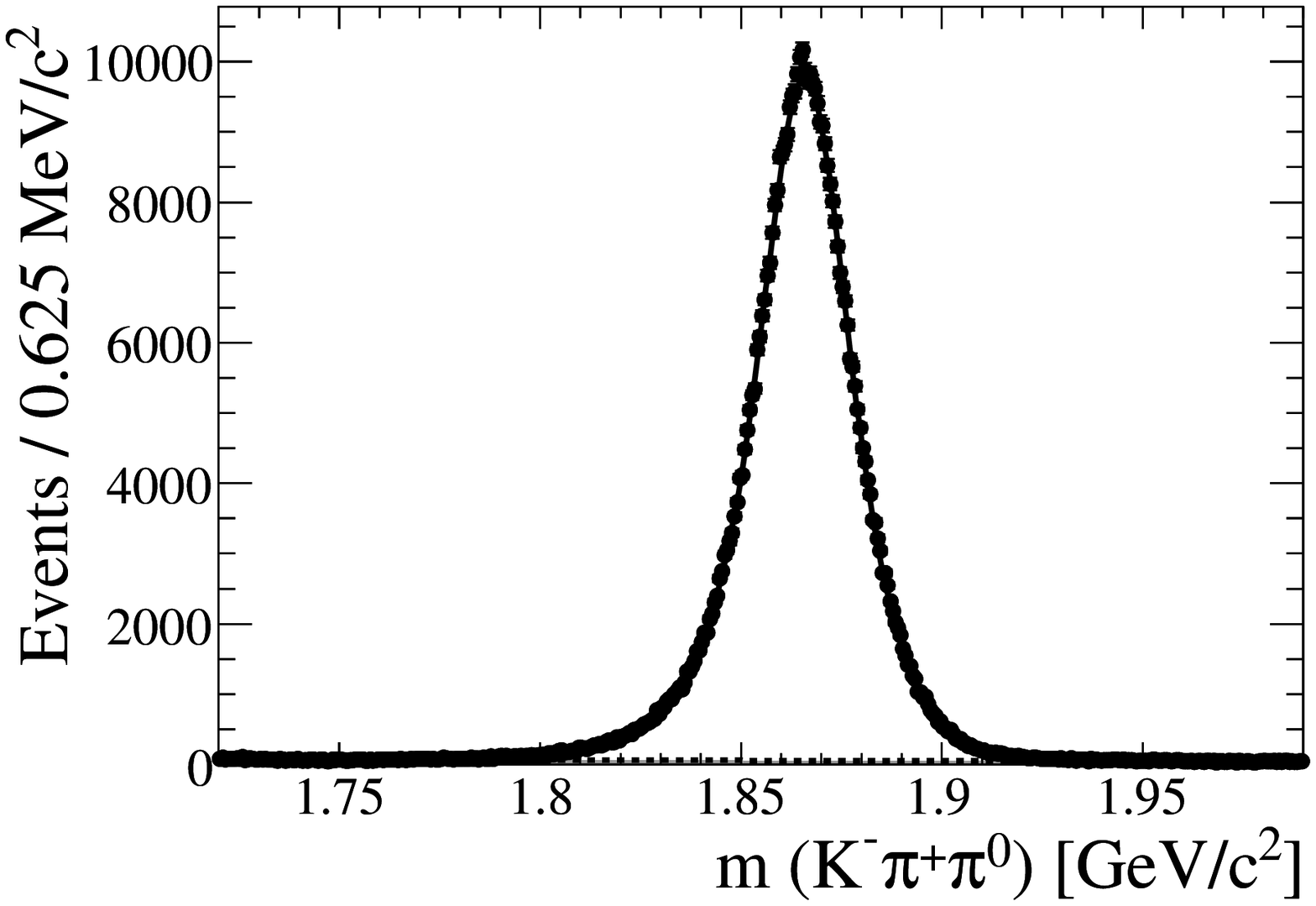} 
\vspace{-0.5em}

\includegraphics[height=1.8in, width=2.8in]{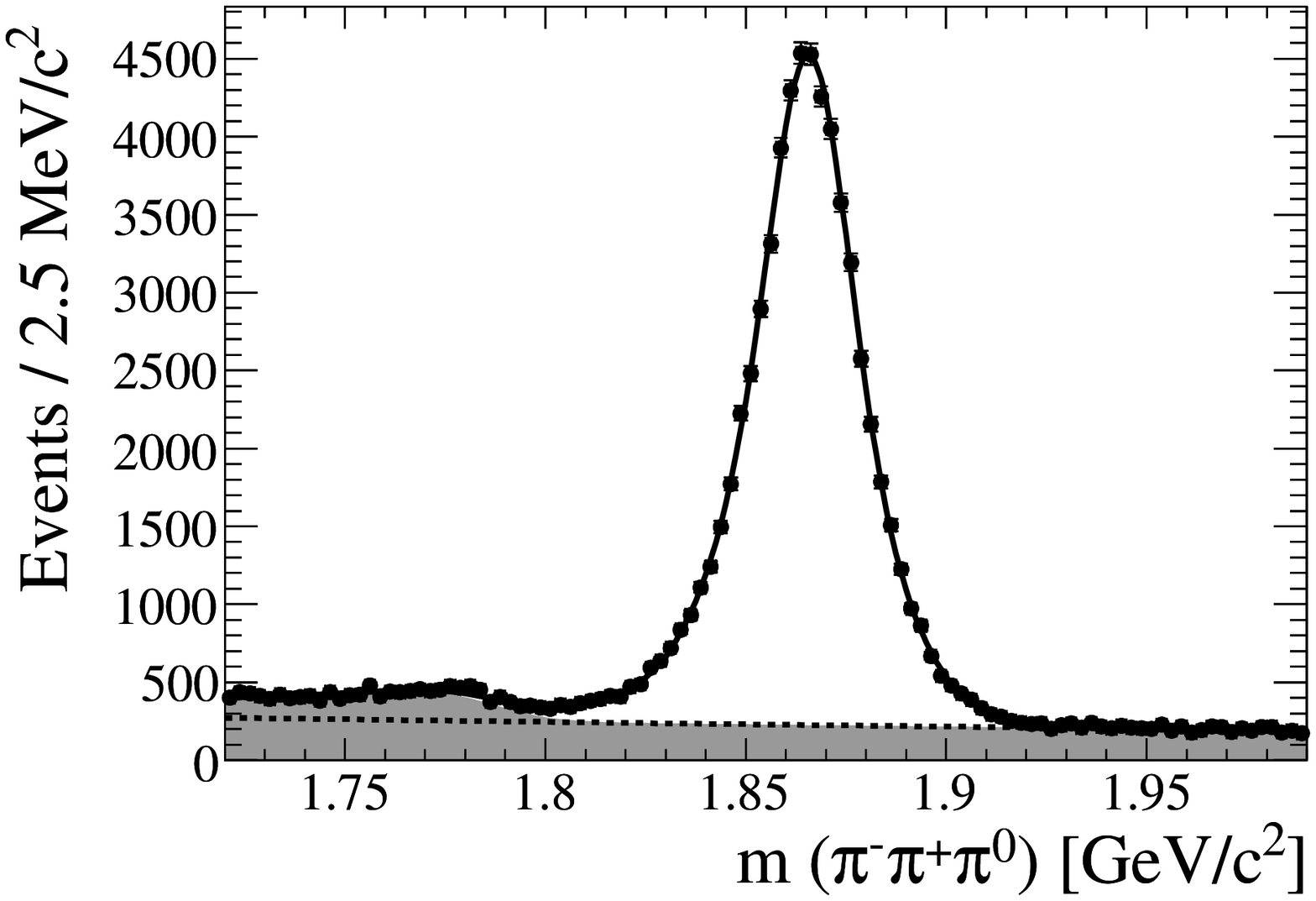} 
\vspace{-0.5em}

\includegraphics[height=1.8in, width=2.8in]{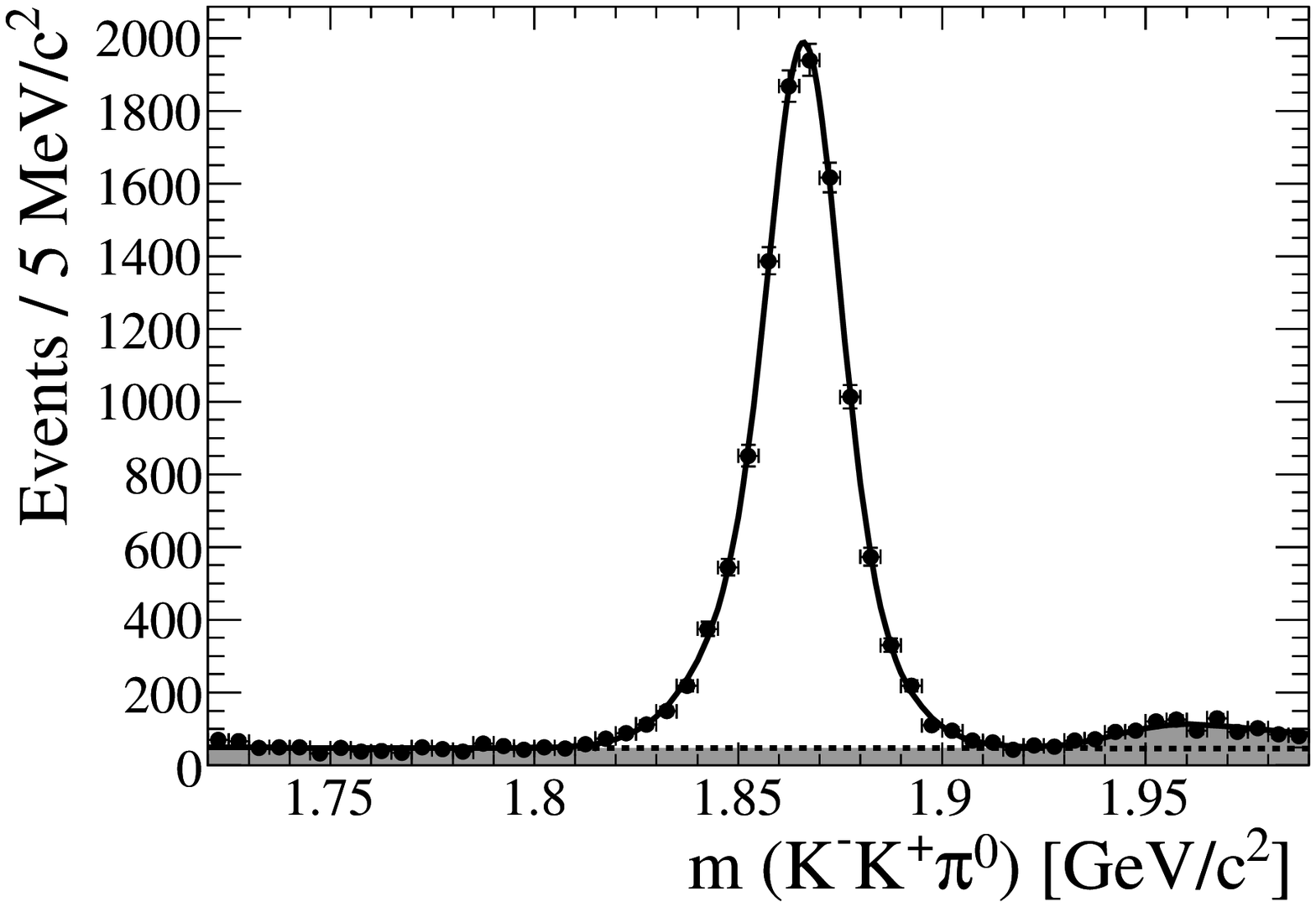} 
\vspace{-1em}
\caption{ Fitted mass for the \kppz, \pppz, and \kkpz\ data samples. Dots are data 
points and the solid curves are the fit. The dot-dashed
lines show the level of combinatorial background in each case. For the \pppz\ and 
the \kkpz\ modes, the shaded region represents the total background. }
\label{FitMassData}
\end{figure}

Cabibbo-suppressed charm decays offer good laboratory for studying weak interactions 
as they provide a unique window on new physics affecting decay-rate dynamics and CP 
violation. The branching ratios of the singly Cabibbo-suppressed decays of \Dz\ meson 
are anomalous since the \Dzpp\ branching fraction is observed to be suppressed 
relative to the \Dzkk\ by a factor of almost three, even though the phase 
space for the former is larger~\cite{pdg2006}.
The branching ratios of the three-body decays~\cite{MarkIII,CLEO} have 
larger uncertainties but do not appear to exhibit the same suppression. 
This motivates the current study which measures the branching ratios of \Dzpppz\ 
and \kmkppz\ with respect to the Cabbibo-favored decay \Dzkppz~\cite{footnote}.\\
\indent This analysis uses a data sample corresponding to an integrated luminosity 
of 232 \invfb\ of \epem\ collisions collected 
around $\sqrt{s} \approx 10$~GeV with the \babar\ detector~\cite{detector} at the \pep2\ 
asymmetric-energy storage rings. Tracking of charged particles is provided by a five-layer 
silicon vertex tracker (SVT) and a 40-layer drift chamber (DCH). Particle
identification (PID) is provided by a likelihood-based algorithm which uses  
ionization energy loss in the DCH and SVT, and 
Cherenkov photons detected in a ring-imaging detector (DIRC). 
A large control sample of $\Dstarp \to \Dz(K^-\pip)\pip_s$ events is used to evaluate 
PID performance for kaons and pions from data. The average identification 
efficiencies and the misidentification rates for pions (kaons) are 95\% (90\%) 
and 1\% (3\%) respectively.   
The typical separation between pions and kaons varies from 8 standard 
deviations ($\sigma$) at momenta of 2 \gevc\ to 2.5$\sigma$ at 4 $\gevc$.
An electromagnetic calorimeter (EMC) is used to identify electrons and photons. 
These systems are mounted inside a 1.5-T solenoidal magnet. The GEANT4 
package~\cite{geant4} is used to simulate the response of the detector 
with varying accelerator and detector conditions. 
Event reconstruction efficiency is obtained using Monte Carlo (MC) 
simulated events having production characteristics from the JETSET~\cite{JETSET} 
fragmentation algorithm.
Three-body $ D^0 $ decays are generated with uniform
Dalitz plot~\cite{Dalitz} (phase space) distributions.
Electromagnetic radiation from  final state charged particles (FSR)
is modeled using the PHOTOS package~\cite{PHOTOS}.\\
\indent To reduce combinatorial backgrounds, we reconstruct \Dz\ candidates in decays 
$\DstarDzpis$ ($\pi_s^+$ is a soft, low momentum charged pion) with $\Dzkppz$, $\pppz$, 
and $\kkpz$, by selecting events with at least three charged tracks and a neutral pion. 
Photon candidates are reconstructed from 
calorimeter energy deposits above 100 MeV, which are not matched to charged tracks.
Neutral pions are reconstructed from pairs of photons with an 
invariant mass in the range 115--160 \mevcc\ and total energy in the laboratory system 
above 350 MeV. 
Charged kaon and pion and $\pi_s^+$ candidate tracks are required to be 
within the fiducial volumes of the tracking 
and PID systems;
they must have at least 20 hits in the DCH 
and transverse momenta greater than 0.1 $\gevc$. 
Also, they must pass PID selection criteria. \\
\indent To form a \Dz\ candidate, two oppositely charged tracks and \piz\ are fit to 
a common vertex, constraining 
the $\gamma\gamma$ invariant mass to the nominal \piz\ mass. The invariant  mass of 
the \Dz\ candidate after the vertex fit is required to lie in the range 1.7--2.0\gevcc. 
To reduce high multiplicity events and combinatorial backgrounds, the momentum of 
the \Dz\ candidate in the event's center-of-mass frame ($p^*$) is required to be 
greater than 2.77 \gevc\ (this requirement also ensures that \Dz\ candidates from B 
decay are removed). The selected candidates after the above requirements are combined 
with the $\pi_s^+$ track to form a \Dstar$^+$ candidate. The \Dz\ and the $\pi_s^+$ are 
constrained to originate from the collison point; the resolution in $\Deltam$, defined 
as the difference in invariant masses of the $D^{*+}$ and \Dz\ candidates, 
is approximately 0.3 \mevcc\ for all three modes.
Only those candidates are retained for which the vertex fit to the whole decay chain 
has a $\chi^2$ probability greater than 0.01 and $\Deltam$ is in the range 
144.9--146.1 $\mevcc$. At this stage, approximately 3\% of the events have multiple 
$\Dstarp\to\Dz\pi_s^+$ candidates satisfying our selection criteria, due to \Dz\ 
misreconstruction or a correctly reconstructed \Dz\ combining with a fake $\pi_s$.
When there is more than one candidate in an event, we select only the candidate with 
the lowest vertex fit $\chi^2$. Our selection procedures result in $\kppz$, $\pppz$, 
and $\kkpz$ samples with purities 99\%, 95\% and 96\%, respectively.\\
\indent The number of \Dz\ signal events in each decay mode is obtained by fitting 
the observed \Dz\ candidate mass distribution to the sum of signal and background 
components, where the latter has combinatorial contributions
and reflection contributions from real three-body
$ D^0 $ decays where a kaon (pion) is misidentifed as a pion (kaon).
The signal component is described by a sum of three Gaussians whose means and widths 
are allowed to vary. 
The combinatorial background is modeled by a linear function.
According to the MC simulation, a large 
fraction of the background consists of $\epem\to\ccbar$ events, with small contributions 
from processes $\epem \to \bbbar$, $u\bar{u}$, $d\bar{d}$, $s\bar{s}$. 
Reflected \kppz\ events peak in the lower (upper) sideband of $m_{\pppz}$ ($m_{\kkpz}$). 
The levels 
of various  background contributions in the \pppz\ and  the \kkpz\ invariant mass distributions 
are shown in Fig.~\ref{generic}. 
The shapes of the \kppz\ reflections in the \pppz\ and \kkpz\ 
invariant mass distributions are obtained from MC. 
The numbers of reflected events are found by making the
\kppz\ invariant mass distributions for the \pppz\ and \kkpz\
samples and fitting them.
Finally, maximum-likelihood fits are performed to extract the signal yields 
from the data for each of the three modes. 
The $\Dz \to K_S^0\piz$ decay is a Cabibbo-favored 
decay and a background for the \Dzpppz\ mode. The level of this  
contamination is obtained by fitting the $K_S^0$ peak in the $m_{\pi^-\pi^+}$ distribution 
and the number of $K_S^0 \piz$ events is subtracted 
from the \pppz\ signal yield. The fitted \Dz\ candidate mass plots for the three modes 
are shown in Fig.~\ref{FitMassData} and the results of the fits are reported in 
Table~\ref{tab:fitresults}.\\
\begin{table}[!htbp]
\begin{tabular}{|c|c|c|c|} \hline 
  Mode     & Number of              & Central value of       & Resolution \\
           & signal events ($S$)    & \Dz\ mass ($\gevcc$)& ($\mevcc$) \\ \hline
  \kppz   & 505660 $\pm$ 750 & 1.8646 $\pm$ 0.0002 & 16.0 $\pm$ 0.5 \\ \hline
  \pppz   & 60426  $\pm$ 343 & 1.8637 $\pm$ 0.0004 & 17.4 $\pm$ 0.8 \\  \hline
  \kkpz   & 10773 $\pm$ 122  & 1.8649 $\pm$ 0.0004 & 13.5 $\pm$ 1.0 \\\hline
\end{tabular}
\caption{ Number of observed signal events and the central value 
and resolution of the \Dz\ candidate mass distribution obtained from fit. The central 
value and the resolution  are, respectively, the average mean and rms width  
of three Gaussians in the signal weighted by their fit-fractions. 
The errors are statistical only.}
\label{tab:fitresults}
\end{table}
\indent 
The event reconstruction efficiency is obtained from MC. The reconstruction 
efficiency for each event is calculated as a function of its position in the 
\Dz\ Dalitz plot. That position is calculated using track momenta
from a fit which constrains the $ h^- h^+ \pi^0 $ invariant mass
to be the nominal $ D^0 $ mass, where `h' is either a kaon or a pion.
To correct for the differences in PID efficiency in 
data and MC, the ratio of these is determined for each track in bins of 
momentum and polar angle, and an event-by-event PID-correction factor is applied to 
each reconstructed event. 
The inverse of the calculated efficiency for each data point 
is taken as its weight. 
The average weight for each decay  mode is computed by summing 
the weights of all events in the  nominal signal regions
($ \pm 3 \sigma$ around the observed mean values of the \Dz\ mass distributions)
and subtracting the efficiency-corrected event 
yields from sidebands 
(1.75--1.79 \gevcc\ and 1.95--1.99 \gevcc, spaced almost
symmetrically around the nominal \Dz\ mass)
to account for background events in the signal region. 
For the 
\kppz\ mode both sidebands are used for this purpose; 
for the \pppz\ (\kkpz) mode only 
the upper (lower) sideband is used because of the  \kppz\ reflection in the other 
sideband. 
The average weights obtained from this method are verified to be unbiased.
The average reconstruction weights for $\kppz$, $\pppz$, and $\kkpz$ modes are 
10.75 $\pm$ 0.02, 9.43 $\pm$ 0.02, and 12.61 $\pm$ 0.05 respectively, where the uncertainty 
is due to MC statistics. \\
\indent The branching ratios are obtained from 
\begin{equation}
\frac{ \B(\Dzpppz)}{ \B(\Dzkppz)} 
= \frac{N_{\pppz}\times W_{\pppz}}{N_{\kppz}\times W_{\kppz}},
\label{eqn:brratio1}
\end{equation}
\begin{equation}
\frac{ \B(\Dzkkpz)}{ \B(\Dzkppz)} 
= \frac{N_{\kkpz}\times W_{\kkpz}}{N_{\kppz}\times W_{\kppz}},
\label{eqn:brratio2}
\end{equation}
\noindent where N and $W$ stand for the number of signal events detected and the 
average weight, respectively. \\
\indent The most important
sources of systematic uncertainties in the branching ratios are 
reported in Table~\ref{tab:syst}. The finite statistics of the MC samples used to 
obtain reconstruction efficiencies contributes a small  uncertainty. The uncertainty 
due to the $\Delta m$ selection is estimated by repeating the analysis with 
different selection windows. The systematic uncertainty due to the background subtraction procedure
is obtained by repeating the analysis using \ccbar\ Monte Carlo data and 
subtracting identifiable ``true'' background events in the signal region. 
The uncertainty caused by Dalitz plot binning effects in the efficiency calculation is estimated 
by varying the bin-size. The effect of the 
modeling of the background probability distribution function on the signal 
yield is studied by repeating the fits to the \Dz\ candidate mass distributions with 
exponential and polynomial combinatorial background models. 
The systematic effect due to 
differences in the $p^*$ distribution in data 
and MC was determined by correcting the reconstruction efficiency obtained from MC by 
the the ratio of $p^*$ distributions in data and MC. Charged-particle identification 
studies in the data lead to small corrections applied to each track in the simulation. 
A large control sample of data and MC is studied separately to determine the residual 
PID uncertainties. 
Uncertainty due to potential differences in charged-particle tracking efficiencies 
in data and MC originating from an imprecise knowledge of different kaon and pion
nuclear interaction cross sections and from the approximations
used in our material model simulation, is conservatively assigned.\\
\begin{table}[!htbp]
\begin{tabular}{|c|c|c|} \hline 
 Systematics & $\frac{ \B(\Dzpppz)}{ \B(\Dzkppz)}$  
& $\frac{ \B(\Dzkkpz)}{ \B(\Dzkppz)}$ \\ \hline \hline 
 MC statistics                   &  0.27\%  & 0.47\%    \\ \hline
 $\Delta m$ selection            &  0.30\%  & 0.90\%    \\ \hline
 Bg. Subtraction                 &  0.60\%  & 0.90\%    \\ \hline
 Efficiency binning              &  0.11\%  & 0.24\%    \\ \hline 
 Bg. PDF model                   &  0.16\%  & 0.13\%    \\ \hline
 $p^*$ difference                &  0.24\%  & 0.02\%    \\ \hline
 PID                             &  0.77\%  & 0.84\%    \\ \hline
 Tracking                        &  0.60\%  & 0.60\%    \\ \hline
 $K_S^0$ Removal                 &  0.07\%  & ---      \\ \hline
 Total                           &  1.25\%  & 1.73\%    \\ \hline
\end{tabular} 
\caption{Summary of systematic uncertainties. The total systematic uncertainty was obtained by adding 
the individual contributions in quadrature.}
\label{tab:syst}
\end{table}
\indent As a consistency check, the analysis is performed separately 
for \Dz\ and \Dzb\ events in different ranges of the 
\Dz\ candidate laboratory momenta to look for systematic variations 
as a function of charge or momentum outside the levels accounted for 
in the estimation of statistical and systematic uncertainties. 
The analysis is repeated for different data run periods and on the  
MC sample treated as data.
As yet another cross-check, the branching ratios are measured 
by directly fitting the efficiency-corrected histograms of the \Dz\ invariant 
mass distributions and then taking the ratio of the yields obtained from the fit. The 
results from all these cross-checks are consistent 
with the results of the main analysis.\\
\indent Using equations~\ref{eqn:brratio1} and ~\ref{eqn:brratio2}, we obtain 
the following results for the branching ratios:\\
\begin{equation}
 \frac{\B(\Dzpppz)}{\B(\Dzkppz)} = ( 10.59 \pm 0.06 \pm 0.13 ) \times 10^{-2}, 
\end{equation}
\begin{equation}
 \frac{\B(\Dzkkpz)}{\B(\Dzkppz)} = ( 2.37 \pm 0.03 \pm 0.04 ) \times 10^{-2},
\end{equation}
\noindent where the errors are statistical and systematic, respectively. 
The previous most precise measurements for these branching ratios are 
$(8.40 \pm 3.11) \times 10^{-2}$ and $(0.95 \pm 0.26) \times 10^{-2}$, respectively~\cite{CLEOc}.
We note that the second result differs significantly from the current world average
value. As we consider events with any level of FSR as parts of our signals,
the ratios we measure correspond to those of the so-called ``bare"
decay rates discussed, for example, in Ref.~\cite{FSR}.
Using the world average value~\cite{pdg2006} for the $\Dzkppz$ branching fraction, we obtain, \\
\begin{equation}
\B(\Dzpppz) = ( 1.493 \pm 0.008 \pm 0.018 \pm 0.053 ) \times 10^{-2},
\end{equation}
\begin{equation}
\B(\Dzkkpz) = ( 0.334 \pm 0.004 \pm 0.006 \pm 0.012 ) \times 10^{-2},
\end{equation}
\noindent where the errors are statistical, systematic, and due to the uncertainty of 
$\B(\Dzkppz)$. \\
\indent The decay rate for each process can be written as:  
\begin{equation}
\Gamma = \int d\Phi |{\cal M}|^2,
\end{equation}
\noindent where $\Gamma$ is the decay rate to a particular three-body final state, 
${\cal M}$ is the decay matrix element, and $ \Phi $ is the phase space. 
Integrating over the Dalitz plot assuming a uniform phase space density, the above 
equation can be written as:
\begin{equation}
\Gamma = \langle | {\cal M} | ^2 \rangle \ \times \Phi,
\end{equation}
\noindent where $ \langle | {\cal M} | ^2 \rangle $ is the average 
value of $ | {\cal M} | ^2 $ over the Dalitz plot and the 
three-body phase space, $\Phi$ is proportional to the area of the Dalitz plot.
For the three signal decays $\Phi$  is in the ratio $\pppz:\kppz:\kkpz$  = 5.05 : 3.19 : 1.67.
Combining the statistical and systematic errors, we find:
\begin{equation}
{ {\langle|{\cal M}|^2\rangle( D^0 \to \pi^- \pi^+ \pi^0 ) } \over
  {\langle|{\cal M}|^2\rangle( D^0 \to K^- \pi^+ \pi^0 ) }   } =  (6.68 \pm 0.04  \pm 0.08)\times 10^{-2} 
\label{eqn:Mratio1}
\end{equation}
\begin{equation}
{ {\langle|{\cal M}|^2\rangle( D^0 \to K^- K^+ \pi^0 ) } \over
  {\langle|{\cal M}|^2\rangle( D^0 \to K^- \pi^+ \pi^0 ) }   } = (4.53 \pm 0.06  \pm 0.08)\times 10^{-2} 
\label{eqn:Mratio2}
\end{equation}
\begin{equation}
{ {\langle|{\cal M}|^2\rangle( D^0 \to K^- K^+ \pi^0 ) } \over
  {\langle|{\cal M}|^2\rangle( D^0 \to \pi^- \pi^+ \pi^0 ) } } = (6.78 \pm 0.14  \pm 0.21)\times 10^{-1}. 
\label{eqn:Mratio3}
\end{equation}
\indent To the extent that the differences in the matrix elements are only due to 
Cabibbo-suppression at the quark level, the ratios of 
the  matrix elements squared for singly Cabibbo-suppressed decays to  that 
for the Cabibbo-favored decay should be approximately 
$ \sin^2 \theta_C \approx 0.05 $ and the ratio of the  matrix elements squared 
for the two singly Cabibbo-suppressed decays should be unity. 
The deviations from this naive picture are less than 35\% for
these three-body decays.
In contrast, the corresponding ratios may be calculated for the
two-body decays $ D^0 \to \pi^- \pi^+ $, $ D^0 \to K^- \pi^+ $,
and $ D^0 \to K^- K^+ $.
Using the world average values for two-body branching 
ratios~\cite{pdg2006}, 
the ratios of the  matrix elements squared for two-body Cabibbo-suppressed decays,
corresponding to Eqs.~\ref{eqn:Mratio1}--\ref{eqn:Mratio3}, 
are, respectively, $0.034 \pm 0.001$, $0.111 \pm 0.002$, and $3.53 \pm 0.12$.
Thus the naive Cabibbo-suppression model works well for three-body decays but not 
so well for two-body decays.\\
\indent In summary, we have measured the  ratios of 
the decay rates for the three-body singly Cabibbo-suppressed decays
\Dzpppz\ and \Dzkkpz\
relative to that for  the Cabibbo-favored decay $\Dzkppz$. 
This constitutes the most precise measurement for these channels to date.
The average squared matrix elements for both of the singly 
Cabibbo-suppressed decays are roughly a factor of $ \sin^2 \theta_C $ 
smaller than that for the Cabibbo-favored decay and are therefore, in 
contrast to the corresponding two-body modes, 
consistent with the naive expectations. \\
\indent We are grateful for the 
extraordinary contributions of our \pep2\ colleagues in
achieving the excellent luminosity and machine conditions
that have made this work possible.
The success of this project also relies critically on the 
expertise and dedication of the computing organizations that 
support \babar.
The collaborating institutions wish to thank 
SLAC for its support and the kind hospitality extended to them. 
This work is supported by the
US Department of Energy
and National Science Foundation, the
Natural Sciences and Engineering Research Council (Canada),
Institute of High Energy Physics (China), the
Commissariat \`a l'Energie Atomique and
Institut National de Physique Nucl\'eaire et de Physique des Particules
(France), the
Bundesministerium f\"ur Bildung und Forschung and
Deutsche Forschungsgemeinschaft
(Germany), the
Istituto Nazionale di Fisica Nucleare (Italy),
the Foundation for Fundamental Research on Matter (The Netherlands),
the Research Council of Norway, the
Ministry of Science and Technology of the Russian Federation, 
Ministerio de Educaci\'on y Ciencia (Spain), and the
Particle Physics and Astronomy Research Council (United Kingdom). 
Individuals have received support from 
the Marie-Curie IEF program (European Union) and
the A. P. Sloan Foundation.
%
%


\begin{thebibliography}{99}
\bibitem{footnote} Reference to the charge-conjugate decays is implied throughout 
the text, unless stated otherwise.
\end{thebibliography}
\end{document}